\documentclass{aa}
\usepackage{txfonts}
\usepackage{graphicx}
\usepackage{natbib}
\bibpunct{(}{)}{;}{a}{}{,} 
\usepackage{umoline}
\def\asec{\ifmmode ^{\prime\prime}\else$^{\prime\prime}$\fi}

\def\it{\sl}
\def\sdss{SDSS~J080434.20$+$510349.2\ }
\def\sdssS{SDSS0804}
\def\degs{\ifmmode ^{\circ}\else$^{\circ}$\fi}
\def\amin{\ifmmode ^{\prime}\else$^{\prime}$\fi}
\def\asec{\ifmmode ^{\prime\prime}\else$^{\prime\prime}$\fi}

\def\degs{\ifmmode ^{\circ}\else$^{\circ}$\fi}
\def\amin{\ifmmode ^{\prime}\else$^{\prime}$\fi}

\def\eqalign#1{\null\,\vcenter{\openup1\jot \m@th
   \ialign{\strut\hfil$\displaystyle{##}$&$\displaystyle{{}##}$\hfil
   \crcr#1\crcr}}\,}
\begin{document}


\authorrunning{Zharikov, S., et al.}
\titlerunning{The accretion disk in the post period-minimum CV SDSS0804 }
\title{The accretion disk in the post period-minimum  \\ cataclysmic variable SDSS~J080434.20$+$510349.2 } 

\author{ S.~Zharikov, G.~Tovmassian, A.~Aviles, R.~Michel,  D.~Gonzalez-Buitrago
\and Ma. T.~Garc\'{i}a-D\'{i}az
}

\institute{
Instituto de Astronom\'{\i}a, Universidad Nacional Autonoma
de M\'exico, Apdo. Postal 877, Ensenada, Baja California, 22800 M\'exico,
}

\offprints{S. Zharikov,\\
\email{zhar@astrosen.unam.mx}}

\date{Received --- ????, accepted --- ????}

\abstract{} {This study of \sdss  is primarily concerned with the double-hump shape in the light curve and its connection with the accretion disk in this bounce-back system. } 
{Time-resolved photometric  and spectroscopic observations were obtained to analyze the behavior of the system between superoutbursts. A qualitative geometric model of a binary system containing a disk with two outer annuli spiral density waves was applied to explain the light curve and the Doppler tomography. }
{Observations were carried out during 2008-2009, after the object's magnitude decreased  to  $V\sim17.7\pm0.1$ from the March 2006 eruption. The light curve clearly shows a sinusoid-like variability with a 0.07 mag amplitude and a 42.48 min periodicity, which is  half of the orbital period of the system. 
In September 2010, the system underwent yet another superoutburst and returned to its quiescent level by the beginning of 2012.  This light curve once again showed a double-hump-shape, but with a significantly smaller ($\sim 0.01$mag) amplitude. Other types of variability like a "mini-outburst" or SDSS1238-like features were not detected.  Doppler tomograms, obtained from spectroscopic data during the same period of time, show a large accretion disk with uneven brightness,  implying the presence of spiral waves.} 
{We constructed  a geometric model of a bounce-back system containing two spiral density waves in the outer annuli of the disk to reproduce the observed light curves.  The Doppler tomograms and the double-hump-shape light curves in quiescence can be explained by a model system containing a massive $\geq $0.7M$_\odot$ white dwarf with a surface temperature of $\sim$12000K, a late-type  brown dwarf, and an accretion disk with two outer annuli spirals.  According to this model, the accretion disk should be  large, extending to the 2:1 resonance radius, and cool ($\sim$2500K). The inner parts of the disk should be optically thin in the continuum or totally void. 
}


{  }

\maketitle

\section{Introduction}

\begin{figure*}[t]
\setlength{\unitlength}{1mm}
\resizebox{11.5cm}{!}{
\begin{picture}(100,60)(0,0)
\put (0,0){\includegraphics[width=16cm, clip=]{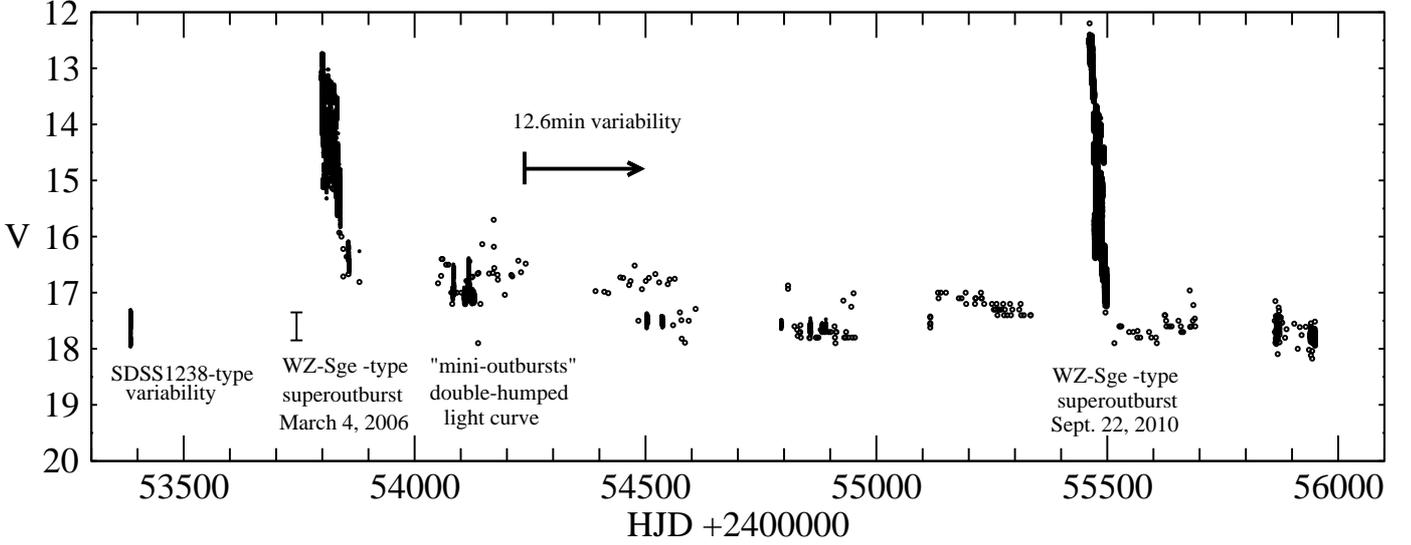}}

\end{picture}}
\caption{Light curves of \sdssS\ are shown from 2005 to 2012 in the $V$ band.  The  plot is based on a combination of data obtained by us, including those published in \citet{2008A&A...486..505Z}, and data accumulated by  the AAVSO.}
\label{lc05-12}
\end{figure*}

A widely accepted evolutionary theory of cataclysmic variables (CVs), as presented in \citet[and references therein]{1999MNRAS.309.1034K}, predicts a significant accumulation of systems around the orbital period minimum \citep{1981AcA....31....1P}.  It also envisions that $\sim$70\% of the current CV population has evolved past the orbital period minimum   and formed   so-called bounce-back systems that  contain a brown dwarf as a secondary \citep{2011MNRAS.411.2695P}. According to the calculations,  the orbital period of a bounce-back system should be slightly longer than before reaching the period minimum.  Bounce-back systems are expected to  float within the 80 - 90 min orbital period range.  However, neither a peak in the period distribution nor any past period-minimum systems  were observed prior to the Sloan Digital Sky Serve (SDSS).  The  SDSS has identified more than a hundred new faint CV systems with  accurately measured orbital periods \citep{2009MNRAS.397.2170G}.  A significant portion of the new CVs have periods clustered at the lower end of the orbital period distribution and some of them are  good bounce-back system candidates.  SDSS J103533.03+055158.4  (hereafter SDSS1035) was proposed by \citet{2008MNRAS.388.1582L} as a solid example of a bounce-back system based upon the system parameters found using a parametrized model of an eclipse observed in its light curve. 
\citet{2010ApJ...711..389A} included SDSS J123813.73-033933.0  (hereafter SDSS1238) 
in the list of bounce-back CVs using a direct IR broad band photometry  of the secondary  and  modeling  the spectral energy distribution of the system.  
 
The subject of this paper  is  SDSS J080434.20+510349.2  (hereafter \sdssS), which has a very peculiar light curve \citep{2006AJ....131..973S}.  A similar  light curve has only been observed in one other system of SDSS1238.  Hence,  SDSS0804 is considered a twin to SDSS1238. Since its discovery, \sdssS\ underwent two superoutbursts, confirming its membership as a WZ\,Sge-type. \citet{2008A&A...486..505Z} included it in the list of   bounce-back candidates based upon its  orbital period and  mass ratio estimates. 
 \sdssS\  exhibits some  phenomena ("brightenings" and "mini-outbursts") that are not  common in other CVs \citep{2008A&A...486..505Z}. Its  superoutburst  in September  2010 \citep{2011arXiv1111.2339P}, which was only about four years after the previous superoutburst, was unexpected. A general description  of  the characteristics and observational features of \sdssS\ can be found in \citet{2006AJ....131..973S}, \citet{2007ASPC..372..511P}, 
 \citet{2008A&A...486..505Z}, \citet{2009PASJ...61..601K}, and \citet{2011arXiv1111.2339P}. 
 
In this paper we report the  results of  monitoring \sdssS\  in 2008-2009 and 2012, 
during periods  when the  object  almost retuned to its  pre-superoutburst, quiescent level. The main aim of this study is to investigate the system's behavior to determine the structure of the accretion disk during the quiescence.

In Sect.~\ref{Obs},  we  describe our observations and the data reduction.
The light curve  analysis and  results of infrared  and spectral observations are presented  in Sects.~\ref{lc} and \ref{ir}.  A discussion of the accretion disk structure of this bounce-back system is given in Sects.~\ref{dop} and \ref{ads}. The light curve simulation of  bounce-back systems is described in Sect.~\ref{lcsim}, and our conclusions are given in  Sect.~\ref{concl}.

\begin{table}
\begin{center}
\caption{Log of the time-resolved photometric observations of SDSS\,J080434.20+510349.2.}
\begin{tabular}{lcclc} \hline \hline
 Date         & HJD start +            & Exp. time                     & Duration \\
              & 2454000                           & num. of integrations          &       \\ \hline
 Photometry   &                                &                               &        \\ 
 2008-02-06 & 502.634       & 30s$\times$821                & 6.8h  \\
 2008-02-07 & 503.626        & 30s$\times$297                & 2.5h  \\
 2008-02-08 & 504.608          & 30s$\times$416                & 3.5h  \\ 
 2008-03-08 & 534.656         & 60s$\times$300                & 5.0h  \\
 2008-03-09 & 535.643        & 60s$\times$225                & 3.8h  \\ 
 2008-03-10 & 536.632          & 60s$\times$250                & 4.2h  \\ 
 2008-03-11 & 537.620         & 60s$\times$290                & 4.8h  \\ 
 2008-11-22 & 593.511         & 120s$\times$90                & 3.0h  \\ 
 2008-11-23 & 594.469         & 120s$\times$41                & 1.4h  \\\hline
 2009-01-23 & 855.724          & 90s$\times$105                & 2.6h  \\ 
 2009-01-24 & 856.690         & 90s$\times$178                & 4.5h  \\ 
 2009-01-25 & 587.688          & 90s$\times$152                & 3.8h  \\ 
 2009-02-27 & 890.705          & 60s$\times$110                & 1.8h  \\ \hline
 2012-01-13 &  1940.786          & 60s$\times$ 107               & 5.6h   \\
 2012-01-17 & 1944.884           &60s$\times$ 87                   & 3.1h   \\
 2012-01-19 & 1946.881           &60s$\times$ 104                 & 3.6h   \\
2012-01-21 & 1948.931           &60s$\times$ 75                   & 2.4h \\
2012-01-22 & 1949.781           &60s$\times$ 149                 & 5.5h \\
2012-01-25 & 1952.780           &60s$\times$ 123                 &5.5h \\
2012-01-26 & 1953.711           &60s$\times$ 179                 &6.7h \\
2012-01-27 & 1954.906           & 60s$\times$ 113                & 2.6h \\
2012-01-28& 1955.836            &60s$\times$ 122                & 4.3h \\
2012-01-29 & 1956.835           & 60s$\times$ 123              & 4.3h \\
2012-01-30& 1957.855            &60s$\times$ 110                & 3.8h \\  \hline
\end{tabular}
\label{tab1}
\end{center}
\end{table}

 \begin{table}
\begin{center}
\caption{Log of the time-resolved spectroscopic  observations of SDSS\,J080434.20+510349.2.}
\begin{tabular}{lcccc} \hline \hline
 Date         & HJD start +            & Exp. time                     & Duration \\
              & 2454000                           & num. of integrations          &       \\ \hline
 Spectroscopy   &                                 &                               &        \\ 
 2008-02-06 & 502.634       & 490s$\times$41                & 8.3h  \\
 2008-02-07 & 503.626        & 490s$\times$53               & 8.0h  \\ \hline
 2008-03-10 & 535.658     & 490s$\times$40                & 5.8h  \\ 
 2008-03-11 & 536.648     & 490s$\times$39               & 5.9h  \\
2008-03-12 & 537.641     & 490s$\times$40                & 6.1h  \\ \hline
2009-01-24 & 855.738     & 483s$\times$32                & 5.3h  \\ 
2009-01-25 & 856.674     & 483s$\times$40              & 6.3h  \\
2009-01-26 & 857.630     & 483s$\times$40                & 6.1h  \\
2009-01-27 & 858.646     & 483s$\times$46                & 6.9h  \\  \hline
2012-01-25 & 1952.670    & 500s$\times$52                & 7.9h  \\
2012-01-26 & 1953.641    & 500s$\times$57               & 8.0h  \\    \hline
 \end{tabular}
\label{tab2}
\end{center}
\end{table}

\section{Observations and data reduction}
\label{Obs} 
 Time-resolved photometry  of SDSS\,0804 was obtained using the direct  CCD image mode
at the 1.5m and 0.84m telescopes of  the Observatorio Astron\'omico Nacional at San Pedro M\'artir
(OAN SPM\footnote{http://www.astrossp.unam.mx}) in Mexico.  The log of photometric observations is given in Table~\ref{tab1}. 
\begin{figure*}[t]
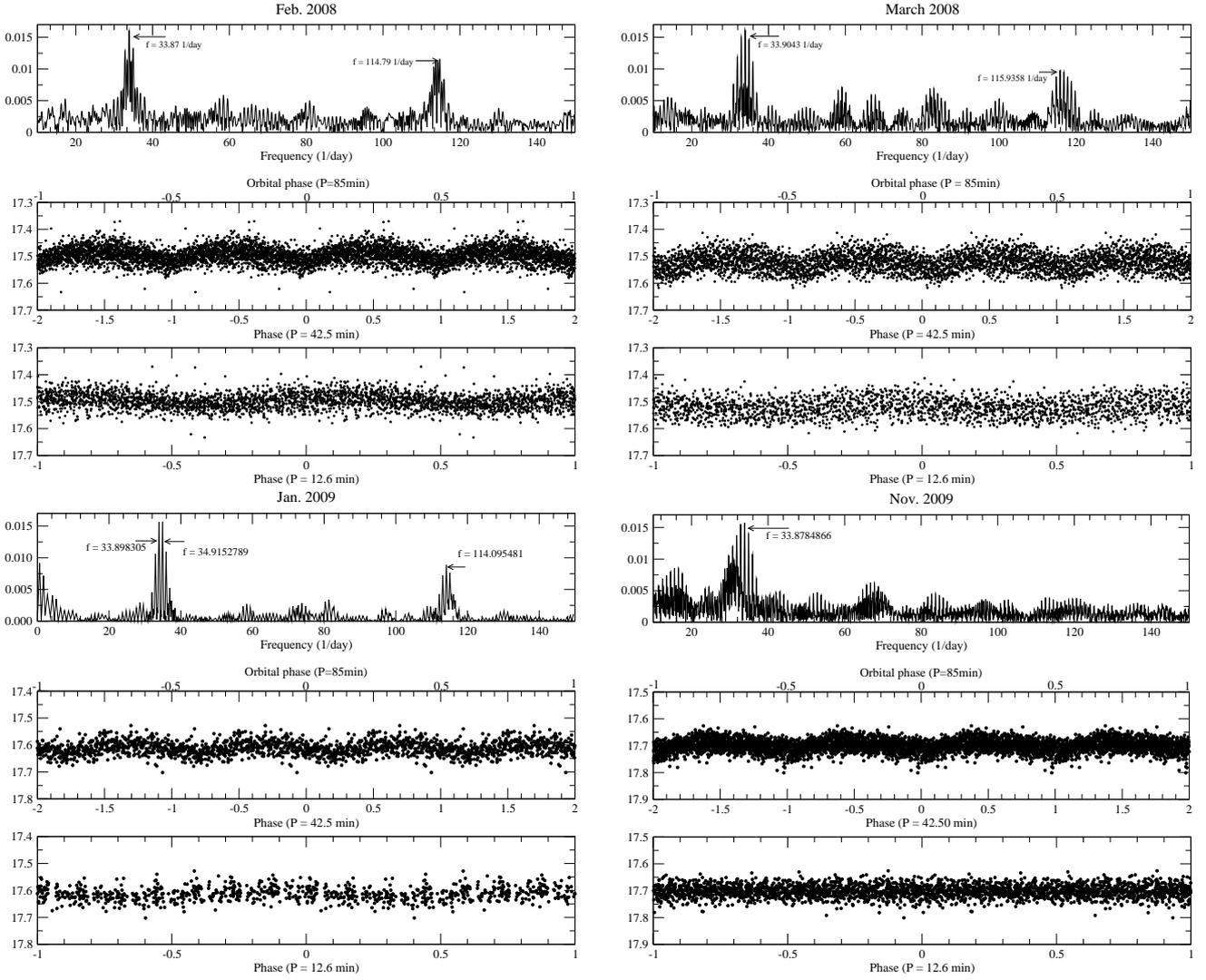

\setlength{\unitlength}{1mm}
\resizebox{12cm}{!}{
\begin{picture}(100,120)(0,0)
\put (0,60){\includegraphics[width=70mm, clip=]{zharikovfig02a.eps}}
\put (75,60){\includegraphics[width=70mm, clip=]{zharikovfig02b.eps}}
\put (0,0){\includegraphics[width=70mm, clip=]{zharikovfig02c.eps}}
\put (75,0){\includegraphics[width=70mm, clip=]{zharikovfig02d.eps}}
\end{picture}}
\caption{Power spectra (top panel) of \sdssS\  is shown for different observational runs; the corresponding  light curves are folded with the orbital period (upper axis of the middle panels). The half orbital period (bottom axis of the middle panels) and the
12.6 min period (lower panels) are also shown. }
\label{folds}
\end{figure*}
Photometric data were calibrated using Landolt's  standard stars. 
 The errors of  photometry were calculated
 from the dispersion of the magnitude of the comparison stars in the object field.  The errors ranged from 0.01 to 0.05 mag.

Long-slit observations were obtained with the Boller \& Chivens spectrograph and the 2.1-m telescope of OAN SPM.  The spectra  span from   
4000 to 7100~\AA\  with a resolution of 3.03~\AA \ pixel$^{-1}$. To improve the signal-to-noise ratio, we obtained a series of phase-locked spectra: ten spectra were taken at equal phase intervals over a single orbital period ($P_{orb} =80.5$\,min). This sequence of spectra was repeated at exactly the same phase intervals for subsequent periods and on subsequent nights. This allowed us to calculate the phase-averaged spectra by summing up spectra at the same orbital phase obtained during several cycles  without  decreasing the time resolution. 
 The log of the spectral observations is given in Table~\ref{tab2}. There is a small difference in the exposure times at different epochs because different   CCDs were used, and each CCD has its own readout time. 
 \begin{figure}[t]
\setlength{\unitlength}{1mm}
\resizebox{12cm}{!}{
\begin{picture}(100,45)(0,0)
\put (0,0){\includegraphics[width=70mm, clip=]{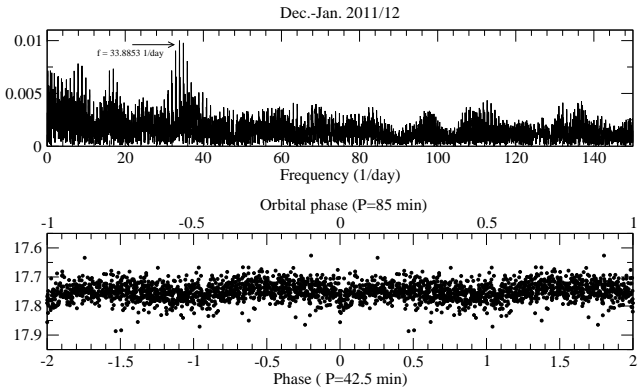}}
\end{picture}}
\caption{Power spectra (top panel) of \sdssS\ using the 2012 data and the corresponding  light curves folded with the orbital period (upper axis of the lower panel) and the half  orbital period (bottom axis of the lower panel) are shown.}
\label{folds1}
\end{figure}

 The $J$ band photometry of \sdssS\  was obtained in October 2009  with the "CAMILA" IR CCD camera 
 attached to the 0.84m telescope of OAN SPM. The data were reduced by the standard way and were calibrated using IR standards observed before and after the object.

\section{Light curve variability}
\label{lc}
The light curve of \sdssS\  during 2005-2012  is shown in Figure\,\ref{lc05-12}. Before the March 2006 superoutburst   the  brightness of the system was $V\sim17.9$  and   the object displayed SDSS1238-like long-term variability (\citet{2006AJ....131..973S, 2006A&A...449..645Z, 2008A&A...486..505Z}). On March 4, 2006,   a  superoutburst commenced, exhibiting all  attributes of WZ Sge-type: a large amplitude  $>$~6 mag, superhumps, and 11 echoes \citep{pavlenko06}. The maximum brightness during the superoutburst reached about $V\sim12$ magnitude.
 Based upon the superhump period \citep{2009PASJ...61..601K}, the mass ratio of the system was  estimated to be $q\approx0.05$ \citep{2008A&A...486..505Z}, which is a typical value for bounce-back systems.

\begin{figure}[t]
\setlength{\unitlength}{1mm}
\resizebox{7cm}{!}{
\begin{picture}(70,55)(0,0)
\put (0,0){\includegraphics[width=8.5cm, bb = 12 220 600 570, clip=]{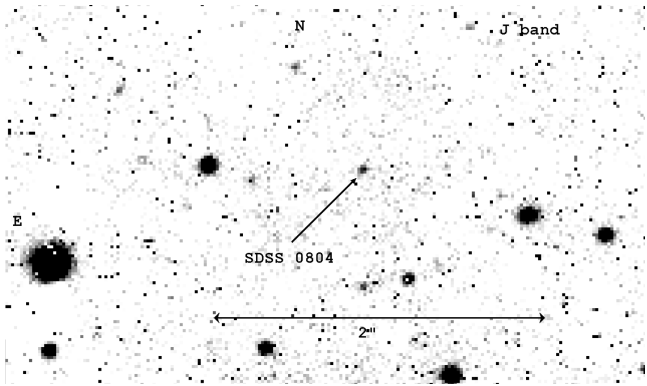}}
\end{picture}}
\caption{$J$ band image obtained in October 2009 using Camila  at the 0.84m telescope of the OAN SPM.}\label{Jband}
\end{figure}

\begin{figure*}[t]
\setlength{\unitlength}{1mm}
\resizebox{11cm}{!}{
\begin{picture}(100,60)(0,0)
\put (10,0){\includegraphics[width=15.5cm, clip=]{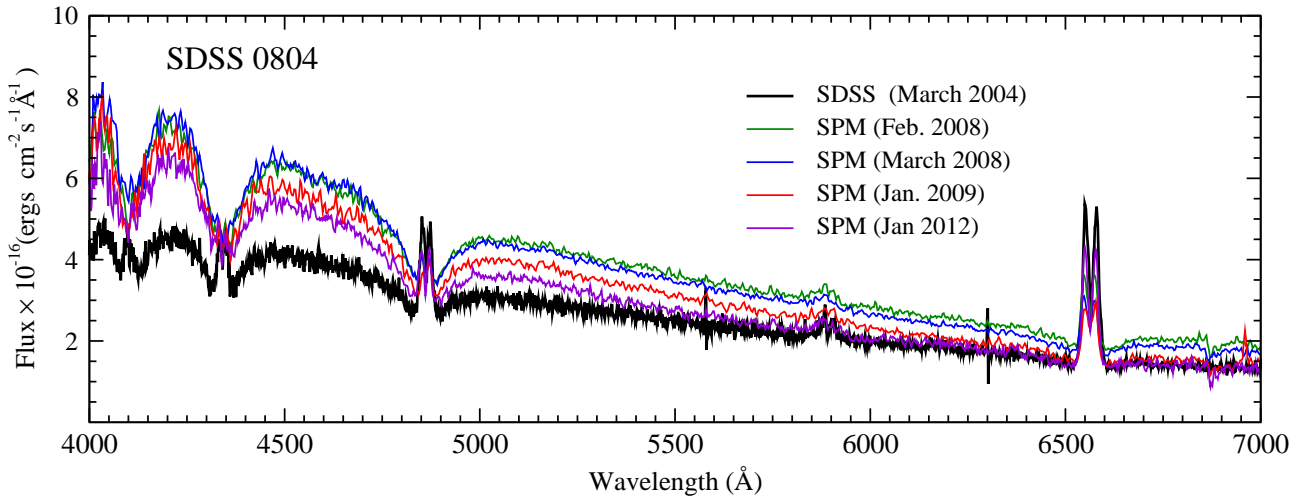}}
\end{picture}}
\caption{Flux-calibrated  time-averaged spectra of \sdssS\  obtained in different epochs.}
\label{spectrum}
\end{figure*}

About eight months  after the first superoutburst, the  brightness decreased to   17.1 magnitude.  Variability at different time-scales  was detected: 
a rapid modulation with a period of $P_1=12.6$ min and  amplitude of $\sim 0.05$ mag was interpreted  as a strong nonradial pulsation of the cooling (i.e., after the outburst) white dwarf, that  entered the instability strip \citep{pavlenko06}; significant short-term variations with periods  $P_2 = 21.7$ min, $P_3 = 14.1$ min, and $P_4 = 4.28$ min that may be additional pulsation modes \citep{2011arXiv1111.2339P}; and the superhumps were replaced with orbital double-hump-shape light variations \citep{2008A&A...486..505Z, 2007ASPC..372..511P}. In addition to these, the object also showed a new phenomenon --  recurring  ($\sim 32$ day) "mini-outbursts"  that lasted approximately four days and increased the brightness to up to $\sim0.6$ mag \citep{2008A&A...486..505Z}. 

Given the diversity of the variability displayed by \sdssS\,, we  continued monitoring  the system with an aim to find  any systematics in its behavior.  In  the three years after the 2006 superoutburst, the V-band magnitude of the object decreased to  about 17.7 mag.  However, the object never
reached the pre-superoutburst level of March 2006 because 
in September 2010 \sdssS\  suddenly underwent another  superoutburst \citep{2011arXiv1111.2339P},   only   $\sim4.5$ years after the  previous one.  This short recess between superoutbursts is atypical  for  WZ Sge-type systems. The amplitude of this superoutburst was similar to the previous, but  the shape of the light curve was different:  the brightness decreased faster and  fewer echoes were detected \citep{2011arXiv1111.2339P}. By December 2011,  more than a year after this superoutburst,  the brightness of the system  fell  practically to the  quiescence level.  From December 2011  to January 2012 we obtained a new set of photometric and spectroscopic  data at quiescence to compare with our previous observations.

The  photometric data were analyzed  for periodicities using a discrete Fourier transformation code \citep{1975Ap&SS..36..137D}. The power spectra for different observational runs are given in Figures\,~\ref{folds} and \ref{folds1}. The power spectra clearly show double  orbital frequency  in all observational runs.  The amplitude of the double-hump-shape light curve decreases with  brightness (from $\sim0.07$\,mag at HJD 2454110,  $\sim0.02$\,mag  at HJD  245456 to only $\sim 0.01$ at HJD 2455953). 
The  12.6 min period, attributed to  WD pulsations, was only detected in the February - March  2008 and January 2009 runs.  This period disappeared afterwards (Figs.\ref{folds} and \ref{folds1}). 
The light curves folded by the 42.5 min  and 12.6 min periods are shown along with their corresponding power spectra in Figures\,\ref{folds} and \ref{folds1} . 
The long-term SDSS1238-like variability and mini-outburst phenomena were not detected  during our monitoring of the system between superoutbursts and after the last superoutburst  (HJD~2454500 --- HJD~2456000). 

\section{Infrared data}
\label{ir}
 \cite{2009PASJ...61..601K} performed infrared photometry of \sdssS\  using OAO/ISLE  in March 2007, 
 about one year after the superoutburst. The IR magnitudes were estimated to be $J = 17.29(5)$, $H=16.97(5)$, and $K = 16.41(6)$. During  these  observations the object was $\sim17.1$ in the visual band,  about 0.7-0.8\,mag above the pre-outburst quiescence level. Our  IR observations were made in the J band in October,  2009. The object  magnitudes was $J=17.3(1)$  (see Fig.\,\ref{Jband}). Hence,  the object did not change significantly in the $J$ band although it was still cooling and becoming fainter in the optical.   We assume that the brightness of the object in the IR, as measured by \citet{2009PASJ...61..601K}, corresponds to the quiescence state.  On  timescales comparable to the orbital period, the system  is expected to show some variability in the 
near-IR  due to the elliptical shape of the Roche-lobe-filling secondary. The ellipsoidal variability of the secondary can be calculated and it has been taken into account as an additional error of about 0.15 mag in the following considerations of spectral energy distribution.

\begin{figure}[t]
\setlength{\unitlength}{1mm}
\resizebox{7.5cm}{!}{
\begin{picture}(70,60)(0,0)
\put (0,0){\includegraphics[width=8.cm, clip=]{zharikovfig06.eps}}
\end{picture}}
\caption{H$_\alpha$ line profile of the object at different epochs is shown. The continuum is subtracted.}
\label{Ha}
\end{figure}

\begin{figure*}[t]
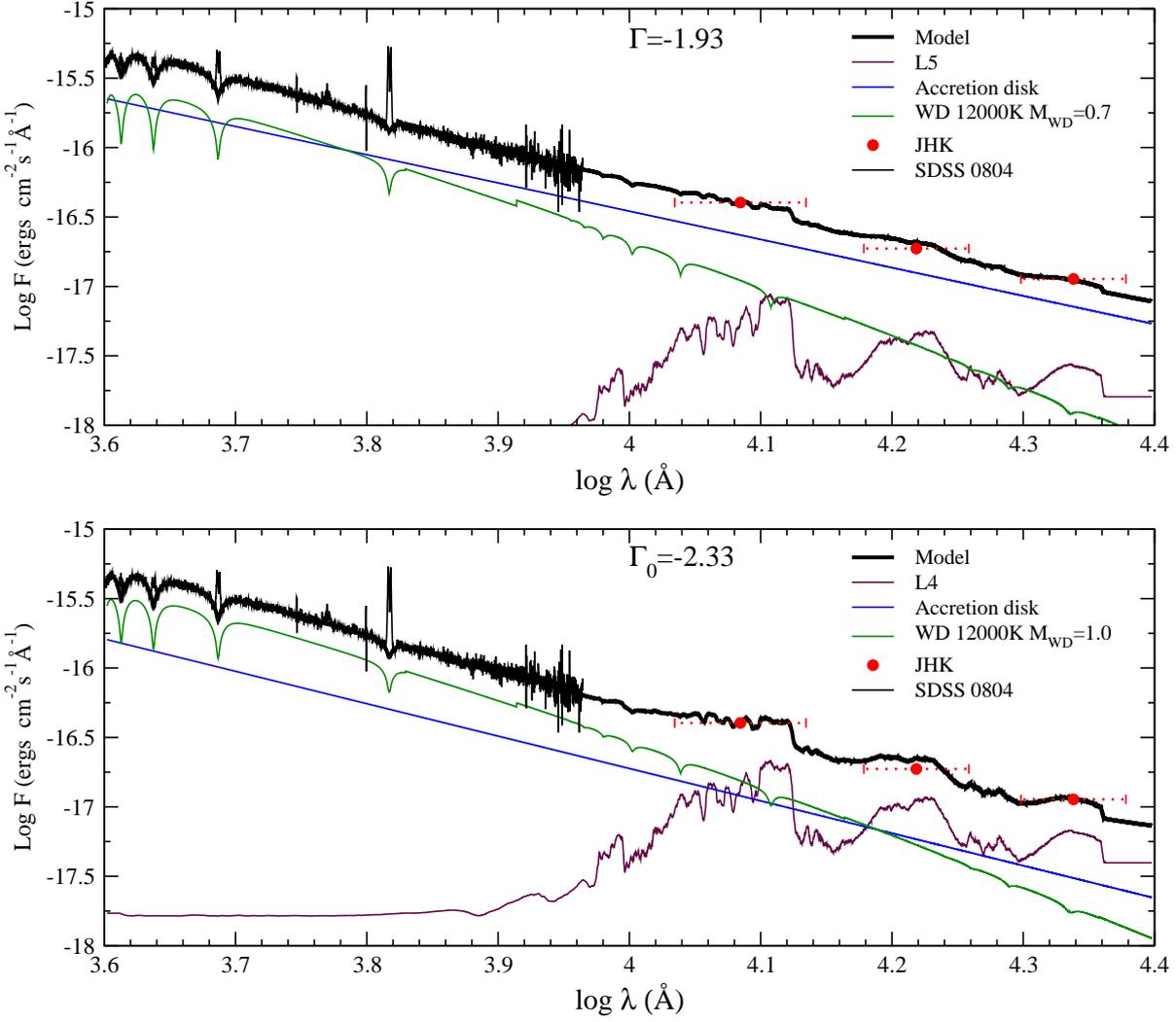

\setlength{\unitlength}{1mm}
\resizebox{11cm}{!}{
\begin{picture}(100,130)(0,0)
\put (0,0){\includegraphics[width=14.5cm, clip=]{zharikovfig07b.eps}}
\put (0,65){\includegraphics[width=14.5cm, clip=]{zharikovfig07a.eps}}
\end{picture}}
\caption{
Spectral energy distribution  of \sdssS\  is plotted with the components of the best-fit models. The tiny black line is the observed quiescent optical spectrum, red dots are $JHK$ IR measurements. Overplotted with a thick black line  is the sum of all components that comprise the close binary.  The green line is the  WD spectral model, the blue line corresponds to the accretion disk contribution, and the maroon line represents the brown dwarf spectrum. The top  figure shows  the solution corresponding  the  $\chi^2$ minimum in which the accretion disk dominates in the IR. The bottom figure  presents the  solution 
with the standard spectral index of $\Gamma = \Gamma_0 \equiv  -7/3 \cong -2.33$ and  the secondary dominating in the IR.
}
\label{fit:spec}
\end{figure*}

\section{Spectral energy distribution} 
\label{sed}

The optical spectrum of \sdssS\ shows a blue continuum with broad absorption lines (caused by the white dwarf) that surrounds the double-peaked Balmer emission lines (caused by the accretion disk). The spectrum undergoes some changes throughout the outburst cycle. The flux-calibrated spectra of SDSS0804, which cover a period of more than two years prior to the  2006  superoutburst until  one year after the  2010 superoutburst,  are shown in the upper panel of Figure\,\ref{spectrum}. 
As already mentioned above, during  the 2008-2009  the object was about $0.3$ mag brighter  than before  the superoutburst. But in January 2012,  the object brightness returned to  the pre-outburst level of March 2006.
The continuum of the post-outburst spectra is bluer, Balmer  absorption lines  are deeper, and  emission lines are weaker than in the pre-superoutburst  spectrum. The equivalent width of H$_\alpha$, for example, changed  from $\approx-90$ to $\approx-15$ in 2004 and 2008.  
In  January  2012, the equivalent width of the H$_\alpha$ line increased to up to $\approx-70$,  about four times larger  than values  obtained between superoutbursts.  However, the value does not reach  the level seen in 2004  (see Figure\,\ref{Ha}) because
the object never reached the same quiescent state  prior to the first superoutburst.

We fit the  spectral energy distribution (SED) of \sdssS\ in quiescence with the model described in detail by  \citet{2010ApJ...711..389A} using a combination of the SDSS  spectrum obtained in  2004  with  the IR data of  \citet{2009PASJ...61..601K}. 
 { Following   \citet{2010ApJ...711..389A}, we assumed that the total  flux,  $F^{*}(\lambda)$, is the sum of the radiation from a  hydrogen WD (DA type) $F_{\rm {WD}}(T_{\rm {eff}}, \lambda)$,   
 an accretion disk with $F_{\rm {AD}} \sim \lambda^{\Gamma}$, 
 and  a red/brown dwarf with $F_{\rm {BD}}(\lambda)$: 
\begin{equation}
 F^{*}(\lambda) = F_{\rm {WD}}(T_{\rm {eff}}, \lambda) + F_{\rm {AD}}(\lambda) +  F^{\rm {SpT}}_{\rm {BD}}(\lambda). 
\end{equation}

\begin{figure}[t]
\setlength{\unitlength}{1mm}
\resizebox{7cm}{!}{
\begin{picture}(70,80)(0,0)
\put (0,0){\includegraphics[width=9.cm, clip=]{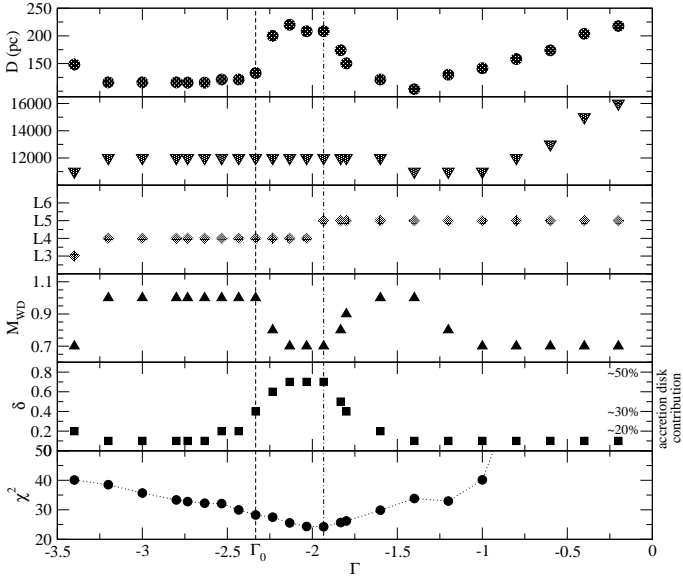}}
\end{picture}}
\caption{Relations of distance, white dwarf temperature, spectral type of the secondary, mass of the primary, contribution of the accretion disk, and $\chi^2$ (from the top to the bottom) vs. the slope of the power law spectrum of the accretion disk. The two models presented in Figure~\protect\ref{fit:spec}  are marked by vertical lines.}
\label{fig:fit}
\end{figure}

Brown dwarf fluxes were  taken from the literature \citep{2003ApJ...596..561M, 2007ApJ...658.1217M} and online sources\footnote{see http://web.mit.edu/ajb/www/browndwarfs/}.
The white dwarf radii were   calculated using the mass-radius  relation  $$R_{\rm {WD}} = 1.12\times 10^9\left(1-\frac{M_{\rm {WD}}}{1.44M_{\odot}}\right)^{\frac{3}{5}}\mathrm{cm}$$ from \citet{1972ApJ...175..417N} 
for an M$_{\rm {WD}} = 0.6 - 1.1$\,M$_\odot$   range, with steps of  a  0.1\,M$_\odot$.  
Spectra of WDs with   pure hydrogen atmosphere in  the 4000-25000\AA\  spectral interval were obtained using ATLAS9  \citep{1993KurCD..13.....K} and SYNTH \citep{1992stma.conf...92P} codes  for the appropriate  range of temperatures.
 The calculations were performed with a 1\,000\,K  step in a temperature range   from  $T_{\rm {eff}}=$10\,000 to 18\,000\,K and with a surface gravity $g= \gamma \frac{M_{\rm {WD}}}{R_{\rm {WD}}^2}$.  The spectra were  normalized to $\lambda_0 = 5500 \mathrm{\AA}$ and the contribution of the WD is 
$$ F_{\rm {WD}}(T_{\rm {eff}}, \lambda) = C_1(\delta) * F^{\rm {norm}}_{\rm {WD}}(T_{\rm {eff}}, \lambda),$$
where $C_1(\delta) = 10^{-0.4*(V+\delta+M^0_V)}$,  V = 17.9 is the object's brightness in quiescence,  and $\delta$ is a parameter that determines the  contribution of the  flux from the  WD to the total flux in the  V band. Finally,  $M^0_V=21.109$  is a constant to convert  magnitude into flux (in $ergs/ cm^2/s / \mathrm{\AA}$)  in the  V band. 
The spectra of the accretion disk  was assumed to be   a simple power law
 $$F_{\rm {AD}}(\lambda) = (C_1(0)- C_1(\delta))\times\left(\frac{\lambda}{\lambda_0}\right)^{\Gamma},$$
where $ (C_1(0)- C_1(\delta))$ determines the contribution from the accretion disk in the V band, assuming that the WD and the accretion disk are the only sources  in that wavelength  because the only other contributor is the brown dwarf,  which  has  an only negligible flux in the V band. $\Gamma$ is  the slope of the accretion disk's power-law spectrum.

 The distance to the object is estimated  to be  $$ d = R_{\rm {WD}} \sqrt{\frac{F^{\rm {bb}}(T_{\rm {eff}}, 5500\mathrm{\AA})}{F_{\rm {WD}}(5500\mathrm{\AA})}}, $$
where $F^{\rm {bb}}(T_{\rm {eff}}; 5500\mathrm{\AA})$ is the black body flux at
$\lambda=5500\mathrm{\AA}$ with an effective temperature $T_{\rm {eff}}$.
Observed SEDs of red/brown dwarfs with spectral types between M6 to L5, normalized to fit the observed flux in  the J band, were used. The bolometric correction to the J magnitude for each spectral type was taken from \cite{Tinney}.

The free parameters of this three-component model are the white dwarf effective temperature $T_{\rm {eff}}$ , the mass of the white dwarf $M_{\rm {WD}}$, the spectral type of the secondary star SpT,  the ratio $\delta$ of WD-to-accretion disk contribution  in the V band, and the slope of the accretion disk spectrum $\Gamma$.

{  Results of the fitting are presented  in Figures\,\ref{fit:spec} and \ref{fig:fit}.  
 \begin{figure*}[t]
\setlength{\unitlength}{1mm}
\resizebox{7.5cm}{!}{
\begin{picture}(70,100)(0,0)
\put (0,100){\includegraphics[width=9.5cm, bb = 145 185 460 750, angle=-90, clip=]{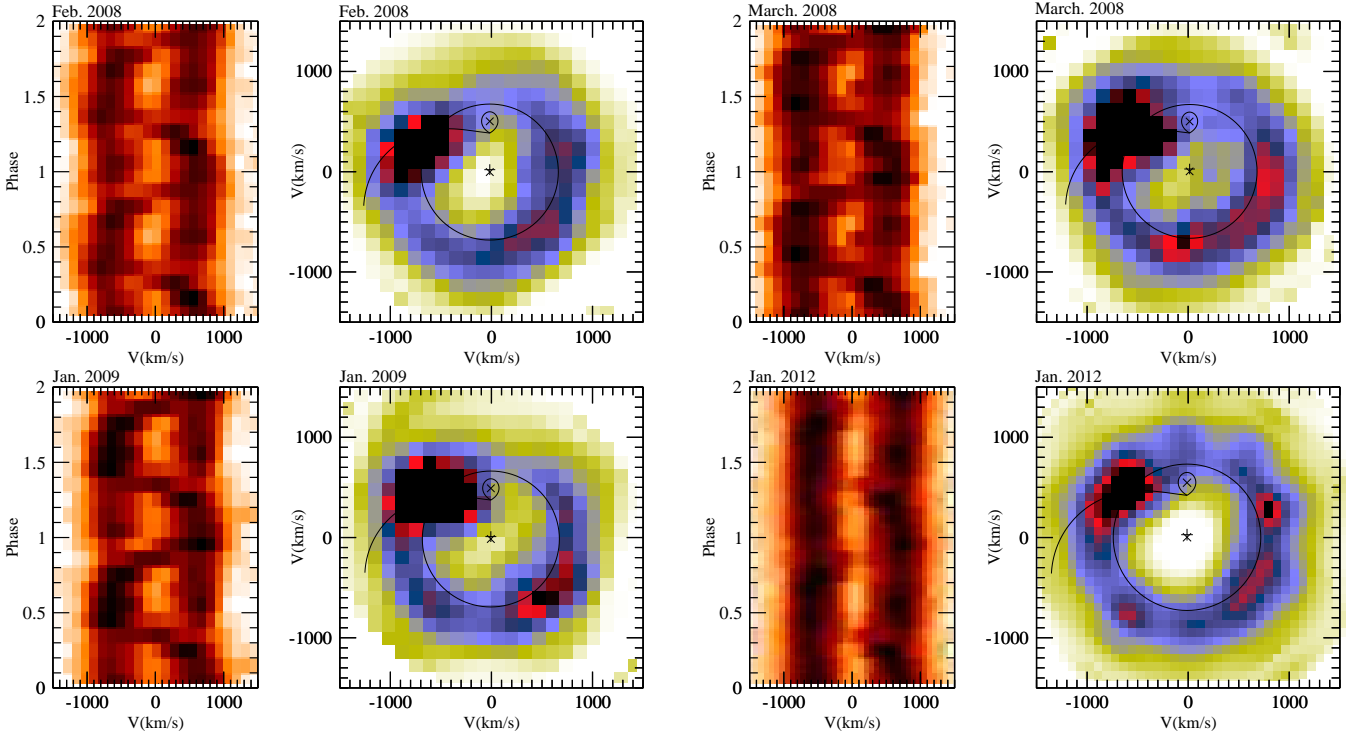}}
\end{picture}}
\caption{Trailed spectra around the H$_\alpha$ line
 folded with the orbital period of the system and the corresponding  Doppler maps.  The center  of  mass, both stellar components, the stream of mass transfer, and the circle corresponding to the 2:1 resonance radius of the accretion disk are marked on the tomograms.}
\label{Dopmap}
\end{figure*}
A marked minimum  in $\chi^2$ occurs at  $\Gamma = -1.93$ (see the vertical dash-dotted  line in Figure\,\ref{fig:fit}).  { This  solution} implies  
 an M$_{\mathrm {WD}}=0.7\pm0.1$M$_{\odot}$ white dwarf with  T$_{\mathrm {WD}}=12000\pm1000$K temperature, a  $\sim$ L$5{\pm2}$ brown dwarf, and an accretion disk  with  a $50$\% contribution in the optical spectrum that dominates  in the infrared. The system will be  located at a distance of about $\sim200$pc.  Indeed, this solution works for a range  of  $-2.2 \lesssim \Gamma  \lesssim -1.8$ where the  $\chi^2$ variation is weak .  
 
 If constraints on $\chi^2$ are relaxed even more, another solution becomes viable for the spectral slope of  
 $ -2.75 \lesssim \Gamma \lesssim -2.2$ and $-1.8 \lesssim \Gamma \lesssim  -1.3$,  which includes the 
 standard spectral index of $\Gamma = \Gamma_0 \equiv  -7/3 \cong -2.33$  \citep{1969Natur.223..690L} 
  (see the dashed line in Figure\,\ref{fig:fit}). The fit parameters for  solutions in these $\Gamma$ ranges   
correspond to  a system formed by a more massive M$_{\mathrm {WD}}=1.0\pm0.1$M$_{\odot}$ white dwarf with the same temperature, a brown dwarf of slightly earlier spectral type  $\sim$ L$4^{+2}_{-3}$, and an accretion disk whose  contribution is  only $\sim30\pm10$\% in the optical spectrum dominated by the white dwarf. In the infrared the contribution of the accretion disk is comparable to  the secondary star. Accordingly, the system would be located at a distance of about $120-130$ pc.

The difference between these two solutions arises from the different contribution of the accretion disk in the infrared.
In the  model with the lowest $\chi^2$  the spectrum  of the accretion disk is flatter compared to the standard model, as expected for an optically thin disk in the continuum \citep{2008NewAR..51..759I}. It also means more contribution from the cooler outskirts of the disk in the infrared. Therefore, if the disk dominates in the IR, the distance estimate to the system increases, which leads to a less massive white dwarf with a larger radius and higher  luminosity. However, there is  evidence that short period CVs tend to contain  massive ($\sim$0.9M$_\odot$) white dwarfs
 \citep{2008MNRAS.388.1582L}. It is difficult to choose the right solution based on quality of the available  data  and the uncertainties  in applied models, but it is beyond doubt that the contribution of the accretion disk in the continuum of bounce-back systems is far less than in ordinary CVs. Additional ultraviolet and infrared spectroscopy would be necessary to determine fractions of fluxes from the stellar components and separate the spectrum of the accretion disk.}

Additionally,  we calculated models for  spectra obtained immediately after the  superoutburst.  The infrared values of the system were kept  the same (within the error range) as in  quiescence.   The temperature inferred for the best-fit model is  T$_{\mathrm {WD}}=15000\pm1000$K.  This value is not  surprising because an  increase of white dwarf surface temperature  during superoutburst   has been observed recently in a number of other CVs \citep{2004ApJ...602..336G, 1996ApJ...471L..41S}. In the case of \sdssS\, the  appearance of WD pulsations after the superoutburst and their subsequent disappearance is also  good evidence of temperature change \citep{2009JPhCS.172a2071P}.

\begin{figure}[t]
\setlength{\unitlength}{1mm}
\resizebox{8cm}{!}{
\begin{picture}(100,65)(0,0)
\put (0,65){\includegraphics[width=63mm, bb = 307 480 460 750, angle=-90, clip=]{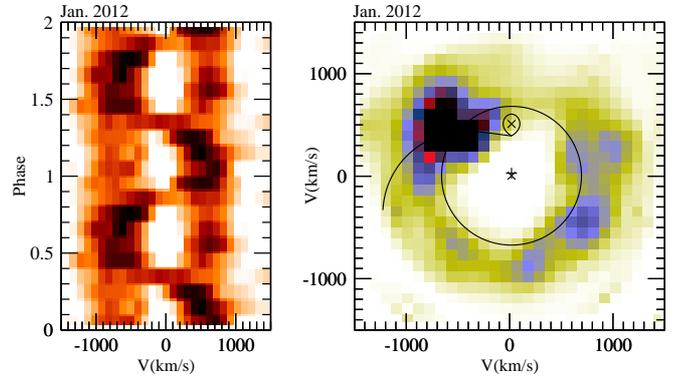}}
\end{picture}}
\caption{Same as shown in Figure 7 but for the H$_\beta$ line and only the 2012 data.}
\label{DopmapHb}
\end{figure}

\begin{figure*}[t]
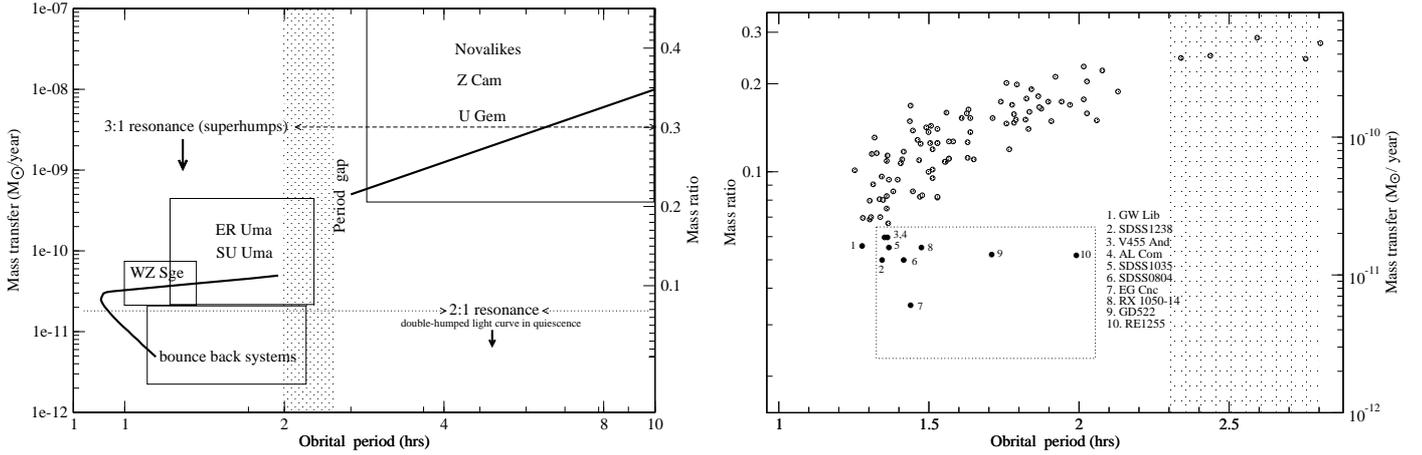

\setlength{\unitlength}{1mm}
\resizebox{10cm}{!}{
\begin{picture}(100,65)(0,0)
\put (0,0){\includegraphics[width=9.2cm, clip=]{zharikovfig11a.eps}}
\put (95,0){\includegraphics[width=9.0cm, clip=]{zharikovfig11b.eps}}
\end{picture}}
\caption{Plots of the  mass ratio and the mass transfer rate vs the orbital period of CVs.  In the schema (left panel), locations of different CV types are marked along with the 3:1 and 2:1 resonance radii. The lower box in that panel corresponds to the location of bounce-back systems, as labeled. In the right panel observed systems with known parameters are plotted with the mass ratio axes marked  on the left and estimated mass transfer rates on the right axes. The bounce-back candidates are enclosed in a dash-dotted box.}
\label{fig8}
\end{figure*}

\begin{table*}[t]
\small
    \caption{Parameters of   bounce-back candidates} 
\begin{center}
\begin{tabular}{l|lcccccccccccl} \hline
NN/Object & P$_{orb}$ &    V       &  q     & M$_1$&   M$_2$ & T$_{eff}^{WD}$ & $i$ & LC$^1$  & References  \\ 
          & (days)   &         (mag)                &         & (M$_\odot$) & (M$_\odot$) & ({\it K}) &($^o$) & &  &\\
  \hline
1. GW Lib$^*$ & 0.0533 & 19.1 & 0.060 & 0.84 & 0.05 & & 11 & &   \citet{2010ApJ...715L.109V}\\ \hline
2. SDSS1238 & 0.056 & 17.8 &     &  1.0 & $<<$0.09 & 12000 & $\ge$70& +q  & Zharikov et al. 2006  \\
3. V455 And$^*$ & 0.0563 & 16.5 & 0.060 &             & $>$M9 &11500 & 83 & +q & {Araujo-Betancor et al.} {2005}\\
4. AL Com$^*$    & 0.0567 &  20.0 & 0.060          &      &    &16300&       & +q       &  \citet{1996PASP..108..748P} \\
5. SDSS1035 & 0.057 & 18.7 & 0.055& 0.94 & 0.05& 10100& 83&                           & \citet{2008MNRAS.388.1582L}\\
6. SDSS0804$^*$ & 0.059 & 17.9 & 0.05  &1.0(1) & 0.05 & 12000 & $\ge$70 & +q & Zharikov et al. 2008\\
7. EG Cnc$^*$ & 0.060 & 18.8 & 0.035 & & & 12300 & & +s                                            &\citet{1998PASP..110.1290P}\\
8. RX1050-14 & 0.062 & 17.6 & $<$0.055& & & 13000 & $<$65&                                     & Mennickent et al. (2001)\\
9. GD552 & 0.0713 & 16.6 & $<$0.052& & $<$0.08& 10900&$<$60 &                             & \citet{2008MNRAS.388..889U}\\
10. RE1255& 0.083& 19.0 & $<$0.06& $>$0.9 & $<$0.08 & 12000 & $<5$ & -                & \citet{2005PASP..117..427P}\\ \hline
11. SDSS1610$^{**}$ & 0.0582 & 19.0 &&&&&& +q & \citet{2010ApJ...714.1702M} \\

\hline
\end{tabular}
\end{center}
$^1$ light curve (LC) features: "+" LC shows  a double-hump during the orbital period; \\ "s" - during  superoutburst; "q" - during quiescence; "-"  absence of double-humps in LC. \\
$^*$ - objects that demonstrate WZ Sge-type superoutburst. \\
$^{**}$ The mass ratio of SDSS1610 is not known, but the object shows similar observational characteristics to selected candidates and the double-humped light curve.
\label{tab.BBS}
\end{table*}

\section{Doppler tomography}
\label{dop}

We used Doppler tomography  \citep{1988MNRAS.235..269M} to study the structure of the accretion disk of \sdssS.  We generated Doppler maps from our time-resolved spectra of the object   using  the maximum
entropy   method as implemented  by \citet{1998astro.ph..6141S}\footnote{ http://www.mpa-garching.mpg.de/$\sim$henk/pub/dopmap/.}.
Trailed spectra around the H$_\alpha$ line and their corresponding  Doppler  maps are shown in Figure\,{\ref{Dopmap}.
The orbital period of P$_{\mathrm {orb}}=0.059$\,days, a white dwarf mass of M$_{\mathrm {WD}}=1.0$\,M$_\odot$, and mass ratio of  $q=0.05$ are used to overlay  positions of  the stellar components and the stream on the Doppler maps\footnote{Using   M$_{\mathrm {WD}}$ = 0.7 M$_\odot$  only  influences overlaid graphics, not the maps; it slightly shifts  the position of the  Roche lobe  and  the accretion stream. The decrease of the circle radius denoting the 2:1 resonance is insignificant (from $\sim$680 to $\sim$\,610km/s).}.  The inclination angle $i=70^o$ is arbitrarily chosen as a compromise  between  strong double-peaked emission lines in the spectra and absence of obvious eclipses in the light curve.  A  possible presence  of weak eclipses during the superoutburst, as reported by \citet{2009PASJ...61..601K}, supports  the idea of a high-inclination system.  Trailed spectra folded with the orbital period of the system clearly show an  S-wave emanating from the bright spot created by the  impact of the stream with the accretion disk.  
The Doppler maps show not only the  bright spot, but also two distinctive bright regions on the ring, indicative of an accretion disk. The bright regions repeat from epoch to epoch. The brighter  one  is located, as  expected,  where the mass transfer stream hits  the accretion disk. This bright region is commonly seen in most CVs, but here it looks extended along the mass transfer stream line.  The other  bright region is extended and is located completely on the opposite side of the disk at $V \approx 750 $\,km/s. This bright region is not  common for accretion disk systems but is a feature shared by the bounce-back candidates  SDSS1238 \citep{2010ApJ...711..389A} and  SDSS1035  \citep{2006MNRAS.373..687S}  in quiescence. 
This latter bright region was also detected in Doppler maps of WZ\,Sge  during its superoutburst in 2001 \citep{2002PASJ...54L...7B, 2003A&A...399..219H}. We assume that the most  probable  interpretation of these tomograms is two spiral waves in the outer annuli, formed by the 2:1 resonance  in the extended accretion disk \citep{1979MNRAS.186..799L}. 

In  January 2012, the intensity of the S-wave in the trailed spectra of the H$_\alpha$ line was  somewhat fainter compared  the total flux from the accretion disk. However, the distribution  of matter in the accretion disk  did not change overall. This observation  is also confirmed by the H$_\beta$ map obtained at the same time (see Figure\,\ref{DopmapHb}).

\section{Accretion disk structure of bounce-back systems}
\label{ads}

Figure~\ref{fig8}  illustrates the current concept of CV evolution  at the orbital period turn-around point and displays the position of the bounce-back candidates on the mass-transfer rate and mass-ratio to orbital period diagrams. Some CV classifications are are labeled on the mass transfer rate to orbital period diagram.  The classification in astronomy is often phenomenological: CVs with orbital periods close to the 80 min orbital period minimum that undergo infrequent (years to decades) superoutbursts are called WZ\,Sge-type stars. Objects with short periods that have not been observed in  outburst or superoutburst but have  spectral characteristics similar to WZ\,Sge (i.e,  a mildly blue continuum with relatively weak  Balmer emission lines  from the accretion disk  embedded in a  broad absorptions formed in the atmosphere of the white dwarf) are listed as WZ\,Sge-type candidates. Bounce-back systems  are spectroscopically  similar to these, but 
 not every WZ\,Sge-type object has necessarily  passed through the turning point. The parameters of bounce-back systems (i.e. objects that would fall into  the  box in  the right panel of Figure\,\ref{fig8}) are listed in  Table~\ref{tab.BBS}.  

To separate the bounce-back systems from the others, the mass of the secondary or the mass ratio is needed. Direct measurement of the mass of either stellar component in CVs is an extremely difficult task. But the mass ratio can be estimated from the period of the superhumps that may be seen in the light curves during superoutbursts.  Four objects (V455 And, AL Com, SDSS0804 and EG Cnc) produced WZ Sge-type superoutbursts in the near past.  Estimates of the  system parameters  often show the presence of  a massive white dwarf  (M$_1\gtrsim 0.9$M$_\odot$) and an extremely low mass ratio  $q\lesssim0.06$, which assumes that the secondary is  a Jupiter-sized brown dwarf  (see Figure\,~\ref{dopmodel}). A $q$ this low implies that the Roche lobe of the primary is huge and the accretion disk could extend to the 2:1 resonance radius and beyond. If that is the case, the disk will develop two spiral density waves in the outer annuli  as shown by \citet{2005ASPC.330.389K}. 
The two waves will cause the appearance of the double hump-shape per orbital period in the light curves, especially in  high-inclination systems.
  Thus, one way of separating  bounce-back systems from other WZ\,Sge objects is to detect the spiral structure in the accretion disk in quiescence. { WZ\,Sge  usually exhibits double-humps only during the early stages of the superoutburst, when the disk  expands.}  In the left panel of Figure\,\ref{fig8}, the mass ratio leading to the 2:1 resonance disk radius is indicated by a tiny dotted line,   which clearly separates bounce-back systems from the rest. 

We aim to determine the conditions in bounce-back systems that  enable detection of these two spiral waves in the outer annuli.
 \begin{figure}[t]
\setlength{\unitlength}{1mm}
\resizebox{10cm}{!}{
\begin{picture}(100,65)(0,0)
\put (10,0){\includegraphics[width=7.cm, clip=]{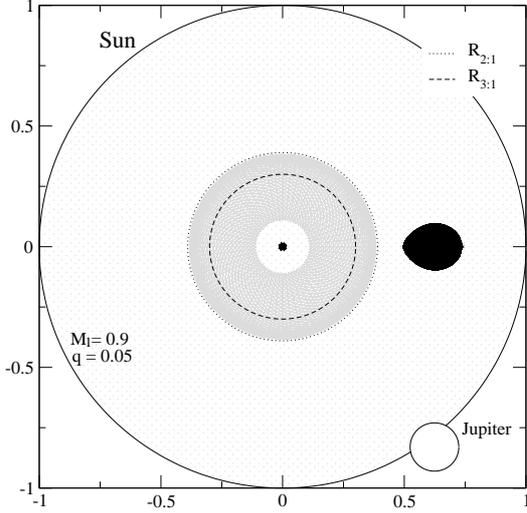}}
\end{picture}}
\caption{
Size of a typical bounce-back system and its components in solar units  compared with the size of Jupiter and the sun. The 2:1 and 3:1 resonance radii are also shown.
}
\label{dopmodel}
\end{figure}
The accretion disk structure of bounce-back candidates is analyzed here primarily based on our spectral and IR photometric observations of SDSS1238 \citep{2010ApJ...711..389A} and \sdssS\ in quiescence. The conditions of the disk are inferred from the Doppler tomography maps 
and from the analysis of  the SEDs of these systems. The SED modeling requires  a low contribution from the accretion disk to the  total flux. The  main uncertainty of the SED model is the slope of the accretion disk continuum.   The shape of a standard spectrum  ($F_\lambda \sim \lambda^{-7/3}$) of the accretion disk is based on a blackbody approximation of a disk element  flux intensity and the relation $T_{eff}(r) \approx T_*(r/R_1)^{-3/4}$  for the radial temperature structure of  a steady-state accretion disk \citep{1995CAS....28.....W}. However,  an optically thick  accretion disk  emitting as a blackbody of this size 
will totally dominate  the  radiation of the system. 
Because observations show that  the white dwarf radiation dominates in the optical range and the secondary's radiation dominates the JHK bands in \sdssS\ and SDSS1238, the "standard accretion disk model" \citep{2002apa..book.....F} does not apply to  bounce-back systems.

 \begin{figure}[t]
\setlength{\unitlength}{1mm}
\resizebox{8cm}{!}{
\begin{picture}(100,80)(0,0)
\put (0,0){\includegraphics[width=11.cm, clip=]{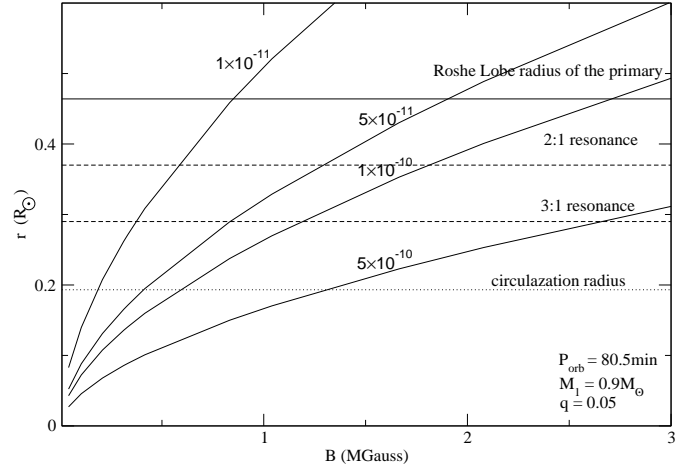}}
\end{picture}}
\caption{ Primary's magnetosphere radius for different mass transfer rates (in  solar mass per year) vs. the primary's magnetic field. }
\label{fDisk}
\end{figure}

 \begin{figure*}[t]
\setlength{\unitlength}{1mm}
\resizebox{12cm}{!}{
\begin{picture}(100,90)(0,0)
\put (0,0){\includegraphics[width=5.5cm,bb = 40 30 600 780,  angle=90, clip=]{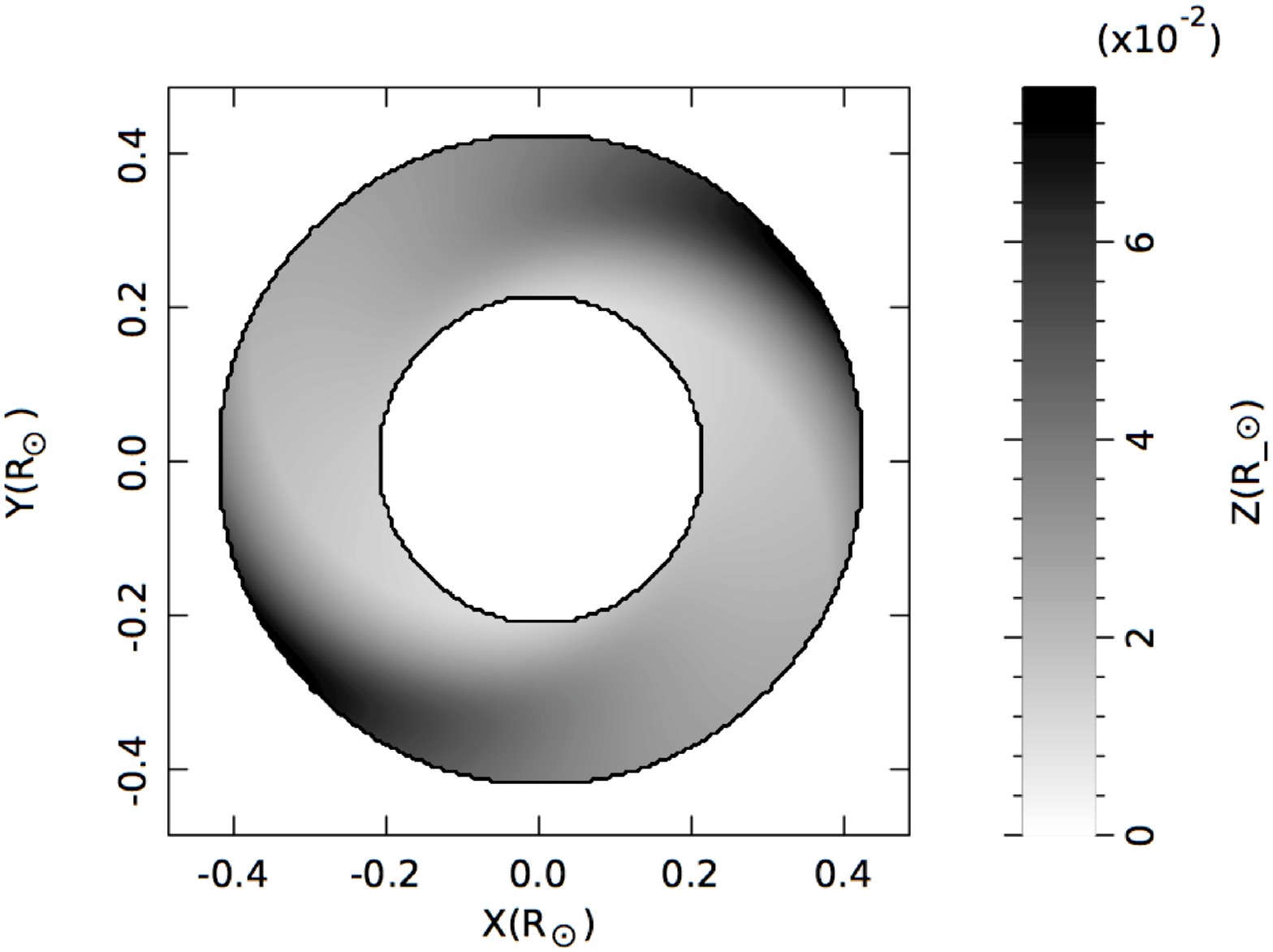}}
\put (75,0){\includegraphics[width=5.3cm,bb =35 10 600 780, angle=90,clip=]{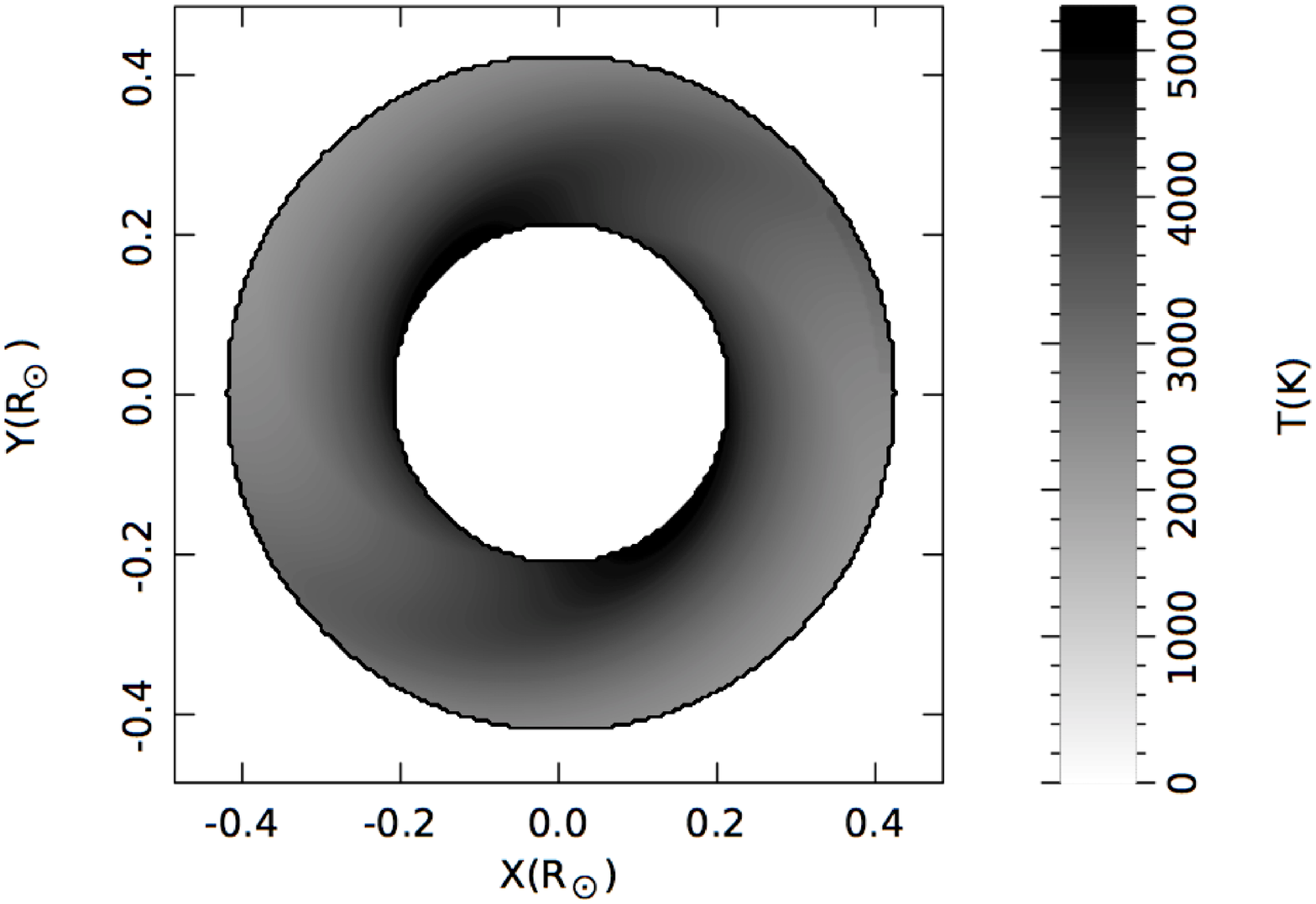}}
\put (30,55){\includegraphics[width=3cm, bb=200 100 400 720,angle=90,   clip=]{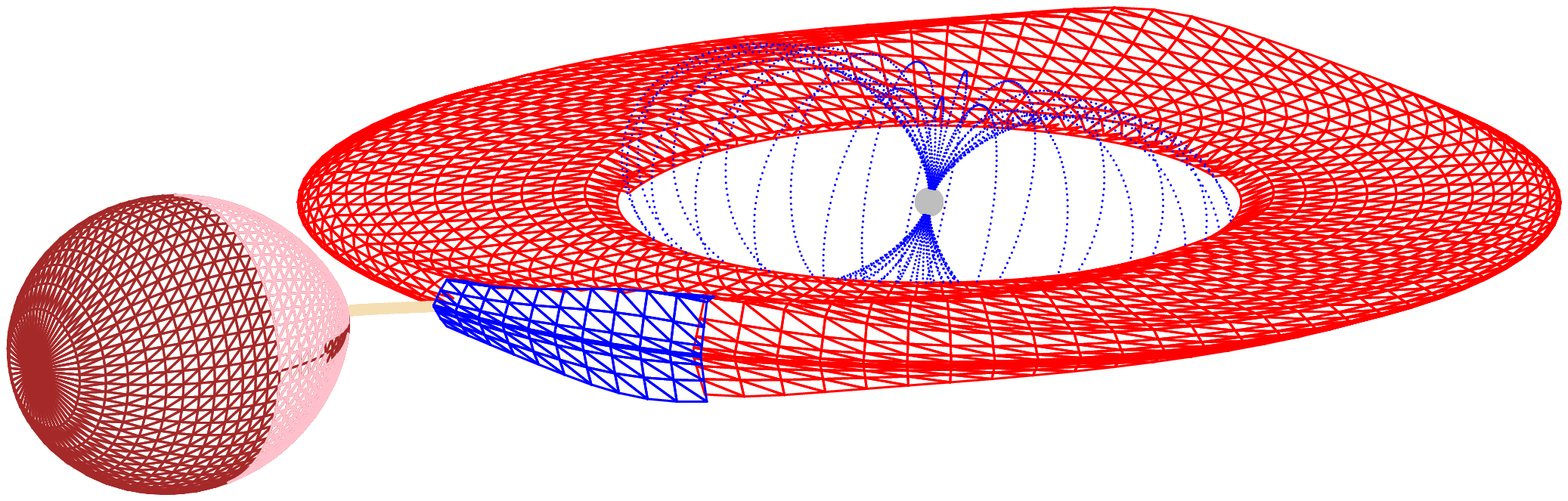}}
\end{picture}}
\caption{Top panel:  model configuration used to calculate the light curves of  bounce-back systems.  The bottom panel shows   the vertical thickness  of the accretion disk (left) and the temperature distribution (right) of the model.
}
\label{model}
\end{figure*}

\begin{figure*}[t]
\setlength{\unitlength}{1mm}
\resizebox{10cm}{!}{
\begin{picture}(100,130)(0,0)
\put (10,3){\includegraphics[width=16.cm, clip=]{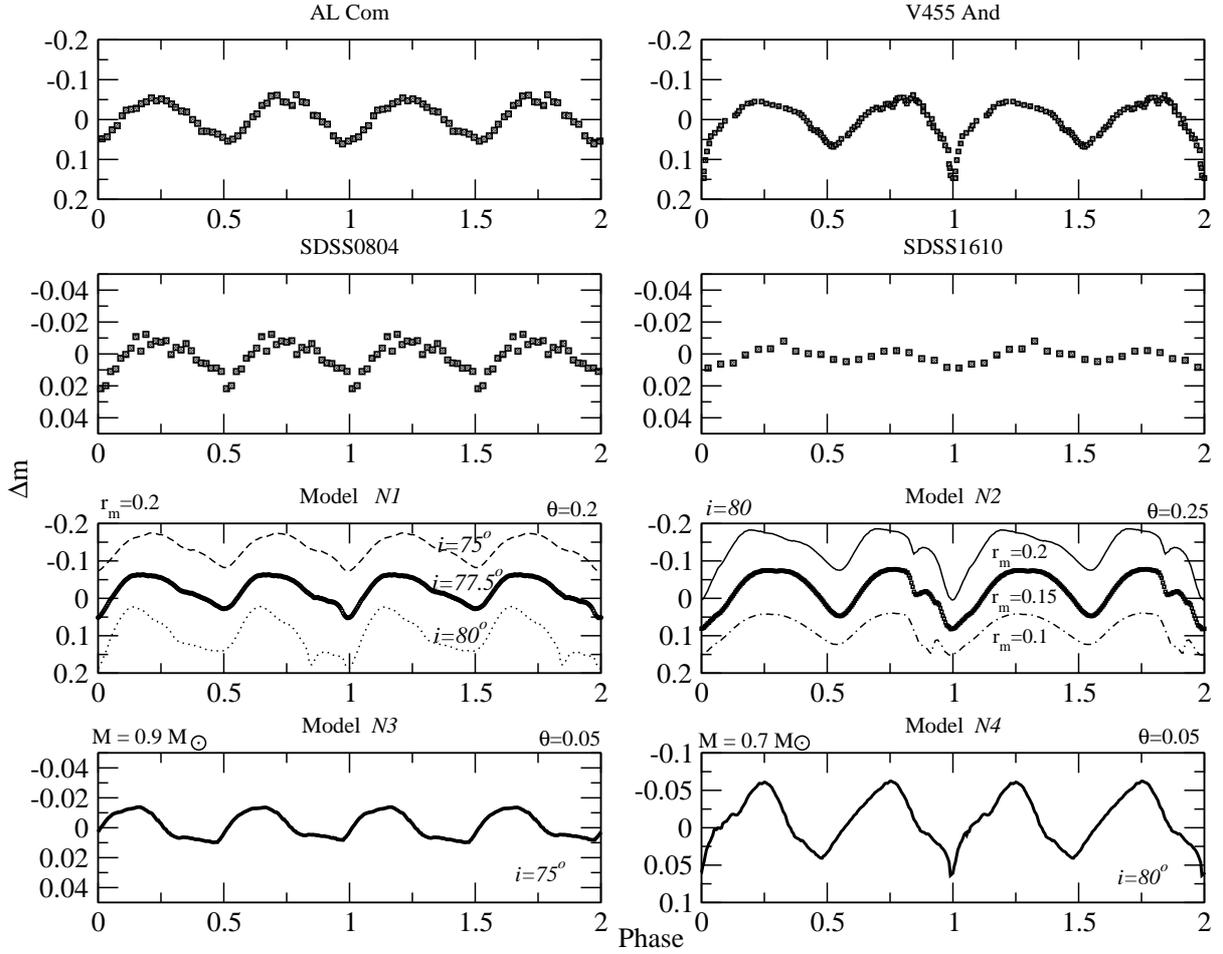}}
\end{picture}}
\caption{ Light curves of AL Com, V455 And, SDSS0804, and SDSS1610 folded with their orbital periods are shown in the upper four panels.  Simulated light curves from our models are shown in the lower four panels (see Table~\ref{tab.Mod}).}
\label{lcmod}
\end{figure*}

It is known that  accretion disks in CVs develop optically thin outer regions for mass transfer rates below about $10^{-8}$ M$_\odot$ yr$^{-1}$  \citep{1980ApJ...235..939W}. \citet{1981AcA....31..127T} confirmed this and added that an increase in  the radius of the $\alpha$-model disk \citep{1973A&A....24..337S} always increases the role of the thin region. For $\alpha \le 1$ the thin part of the disk is rather cool ($\le 6000$K). Because a large part of the accretion disk is optical thin in continuum (i.e., in the non-LTE regime and for $\alpha<0.1$), the disk temperature can drop below 5000K \citep{1991A&A...242..503D}. \citet{1984ApJS...55..367C} studied  the vertical structure of a steady-state, $\alpha$-model thin-accretion disk for an accreting object of 1\,M$_\odot$. They found that, for low accretion rates, the disk structure  is optically thin. For $0.01< \alpha < 1$,  the solution of disk equations (1)-(3) \citep{1984ApJS...55..367C} can be double-valued with high- ($\sim 5000
 $\,K) and low- ($\sim 2000$\,K) temperature branches. For $\alpha> 0.1$ a warm solution is possible in the inner region of the accretion disk, but disk annuli at larger radii will be in a cold state with $T < 2000K$.  Only the low-temperature solution exists for $\alpha\approx 0.1$.  As $\alpha$ decreases with temperature, the tendency to develop cold solutions in quiescence is enhanced. Until now, models of disk emission spectra  for these cool disks have not been calculated,   the nearest approach is a flat spectrum in the range 3000-10000\AA \ for  the   model of the  cool (T$=5000$\,K, $\alpha=0.03$) accretion disk  \citep{2008NewAR..51..759I}.  
The next important aspect regarding accretion disks of bounce-backers  is the non-uniform structure inferred  from the Doppler tomograms and optical light curves alike.  As already mentioned,  the non-uniform  structure was interpreted as  spiral arms in the accretion disks \citep[see Doppler maps discussion and references therein]{2010ApJ...711..389A}.  
The inner part of the accretion disk in a standard model  is usually optically thick and forms a continuum of the disk spectrum \citep[and references therein]{1984ApJS...55..367C}.  However, for WZ-Sge-type systems, various authors have  supported  the idea that the inner part of the disk  is void  during quiescence (see \cite{2011A&A...528A.152K, 2001A&A...376..448M} and reference therein). This void is assumed   to explain the long recurrence time for superoutbursts and the absence of normal outbursts.  
The missing inner part of the accretion disk may be caused by evaporation \citep{1994A&A...288..175M} or/and  by   magnetic field of the primary WD  \citep{2006MNRAS.372.1593M}. According to the equation $ r_{\mathrm m} = (GM_{\mathrm {WD}})^{-\frac{1}{7}}\dot{M}^{-\frac{2}{7}}\mu^\frac{4}{7}, $ where $\mu \equiv  B_{0}R_{\mathrm {WD}}^3\sim10^{32}\ \mathrm{Gauss\ cm^3}$  \citep[and references therein]{2007ApJ...671.1990K}, the 
size of the  magnetosphere will increase with the decrease of  mass transfer rate (see Figure\,\ref{fDisk}). A relatively faint ($\le 1$ MGauss) magnetic field is enough to form  a cavity in the inner part of the accretion disk. 
Magnetic fields of about 0.5 MG and stronger are detected in magnetic CVs from the white dwarf spin modulation observed in optical light or X-rays. V455\,And is a proven intermediate polar.  In \sdssS\, a number of high-frequency periods were detected \citep{2011arXiv1111.2339P}, but none was identified as a spin period so far. X-ray observations are needed to probe the magnetic properties of the white dwarfs in bounce-back candidates to verify the intermediate polar hypothesis.  A typical system with parameters corresponding to a bounce-back system may look like the sketch in Figure\,\ref{model} (top panel).   
The simulated disk is thin in the inner regions and thick along the spiral density waves, which are located in the outer annuli.  

 \begin{table}[t]
\small
    \caption{Parameters  used for  V- band  light curve modeling} 
\begin{center}
\begin{tabular}{l|lcc|cccc} \hline
$N_{mod}$  & $i$       &  r$_m$        &  r$_{out}$ & $\xi^*$ & $\epsilon^*$ & $\delta^*$  & $\theta$ \\
         &  ($^o$) &  R$_\odot$ & R$_\odot$ &                      &           &                      & ($^o$)      \\\hline
N1     & 75-80 &0.2 & 0.45 & 0.7 & 0.06 & 35 & 0.2   \\ 
N2     & 80   &  0.1-0.2  & 0.45 &  0.9 & 0.08 & 25 & 0.25 \\   \hline
N3     & 75   &  0.2  &  0.45 &  0.7 & 0.06 & 35 & 0.05   \\ 
N4$^\dagger$     & 80  &  0.2  &  0.45 &  0.7 & 0.06 & 35 & 0.05 \\   \hline

\end{tabular}
\end{center}
\begin{tabular}{l}
M$_1$ = 0.9M$_\odot$, $q$ = 0.05, T$_1$ = 11000K, T$_2$=2000K, \\ P$_{orb}=5100s = 1.42h$, 
 $a$ = 0.64R$_\odot$, R$_{L1}$ = 0.49R$_\odot$, \\ $\dot{M} = 1.6\times10^{-11}\ M_\odot / yr$, 
 $\beta^*$ = 0.01 \\
$^*$ --- parameters of the disk and the spiral arm corresponding  to \\ the model
 of the two-armed spiral structure in the accretion disk \\ of \citet{2004ApJ...606L.139H} 
 (see equations 3,4 and 5). \\
 $^\dagger$ --- the model with M$_1$ = 0.7M$_\odot$ \\
 $i$  is the inclination; M$_1$ is the mass of primary; $q$ is the mass ratio \\
T$_1$ and T$_2$ are the temperatures of the primary and the secondary \\
$a$ is the system separation;  R$_{L1}$ is the radius of the primary Roshe\\  lobe 
r$_m$ and r$_{out}$ are the inner and outer radii  of the disk

\end{tabular}
\label{tab.Mod}
\end{table}

\section{Light curve simulation}
\label{lcsim}

An eye-catching double-hump-shape per orbital period  in quiescent light curves is a  common feature of objects proposed to be bounce-back candidates. 
We developed a  geometric  model of CV disks with two spiral density waves in the outer annuli of a thin accretion disk  to model  light curves of  bounce-back systems. The top panel of Figure\,\ref{model} presents a snapshot of this model. The system is comprised of a primary white dwarf, a secondary brown dwarf star, a stream of accretion matter, and an accretion disk.  
The white dwarf is a sphere defined by the mass-radii relation (2.83b) from \citet{1995CAS....28.....W}. The secondary is assumed to fill its Roche lobe, and the Roche lobe shape is calculated directly using   equation (2.2) from \citet{1995CAS....28.....W} for equipotential $\Phi$(L$_1)$.
The temperature of the primary and the secondary are defined as initial parameters. Although illumination of the secondary by the primary in these systems
is low, this illumination is also included. A moderate  temperature gradient $T(r) \sim T(r_{in}) \times r^{-3/4}$ between the inner and outer edges of the disk and  $T\sim T(r)\times(1+\theta\times z(r)$) from the spiral density waves (see Figure\,\ref{model}) was assumed.  Here, $T(r_{in})$ is the temperature of the  internal radius of the disk, which depends on the mass ratio (see equations 2.35 and 2.36 in \cite{1995CAS....28.....W}), and $\theta$ is a free parameter that defines the  temperature gradient  in the spiral density waves.  The model  takes into  account the positions of the bright structures in the Doppler maps, the large size of  accretion disk, and the description of the spiral density waves in \citet{2004ApJ...606L.139H}.  Figure\,\ref{model} presents the geometry used in the model (top panel) and  grayscale images present  the height of the disk  (bottom left) and the temperature distribution (bottom right) of the accretion disk.

We calculated a variety of models using the typical average parameters of bounce-back systems presented in   Table \ref{tab.BBS}.  In Figure\,\ref{lcmod} four representative models 
corresponding to the entries of Table\,\ref{tab.Mod} are plotted along  with the observed, averaged   light curves of AL Com, \sdssS, SDSS1610, and V455 And.  
The main difference between the  models is the choice of parameters, which influences the spiral density wave structures  in the disk. 
In the second model (N2), the spiral density waves are hotter and thus the amplitude is higher than in the model N1 (Table\,\ref{tab.Mod}). 
To show  the dependence on system inclination from the orbital plane and the radius of inner disk in the models, we plot  three  curves for each model in  the lower panels of Figure\,\ref{lcmod}.  Obviously, the inclination varies  in the N1 model and the inner disk radius varies in the N2 model.
 For models N3 and N4, a significantly smaller  $\theta$ is used  for the 
$M_{WD}=0.9M_\odot$, $i=75^o$ and $M_{WD}=0.7M_\odot$, $i = 80^o$ cases, respectively.  Clearly, the double-hump-shape light curve is easily reproduced by these models. 
From our models, we find that the shape and  amplitude of the variability depends on the parameters chosen for the two spiral density waves, the disk temperature distribution, the size of the inner cavity, and the outer radius of the disk.  We also find that the amplitude of the double-humped variability decreases with decreasing temperature gradient in the spiral density waves  and/or the decline of the  system's inclination. We note that changing the primary's mass has no effect on the shape of the light curve.  However, as noted previously , a low-mass  primary will not allow the  disk to reach the 2:1 resonance radius, which results in the  spiral structures in the outer annuli of the accretion disk. 

The number of free parameters of the light curve model is large.  Using  a  $\chi^2$ minimization to fit  each light curve precisely would be  a very complicated task and is not the aim of this study. The simple models used in this work validate  previously reported ideas on the disk structure in WZ\,Sge systems - a large, cool, mostly optically thin accretion disk containing  two dense/hot spirals.  We add that these models offer a reasonable explanation for the observed characteristics of bounce-back systems.

\section{Conclusion}
\label{concl}

 We here continued our study of SDSS J080434.20+510349.2  focusing   on  the nature of  double humps in the light curve and their relation to the structure of the accretion disk. 

 The object  exhibited $\approx$0.07 mag variability with   42.48\,min period,   exactly  half   the orbital period of the system, during 2008-2009  when it almost retuned  to  quiescent  level  after the March 2006 superoutburst.   In September 2010, the system underwent yet another superoutburst and approached   its quiescent level by the beginning of 2012. The light curve once again showed  double humps, but with a significantly smaller ($\sim$0.01mag) amplitude.
 No long-term SDSS1238-like variability and Ómini-outburstÓ phenomena were  detected.
 
 The  SED of \sdssS\ in quiescence was fitted by the three-component  model, which 
 requires a massive $\geq$0.7M$_\odot$ white dwarf with a surface temperature of $\sim$12000K, a late-type brown dwarf, and an   accretion disk with a weak contribution to the total flux.
  
H$_\alpha$ and H$_\beta$  Doppler maps obtained in quiescence from  time-resolved spectroscopy  persistently show  excess emission  from two opposing  sides of the accretion disk. This pattern is a common characteristic that \sdssS\ shares with SDSS1238  and other bounce-back candidates.

We constructed a geometric model of a bounce-back system  to reproduce the observed light curves. The  double-humped  light curves in quiescence can be generated by the  model in which the accretion disk  extends to the 2:1 resonance radius, is cool ($\sim$2500K) and contains two spiral density waves in the outer annuli of the disk. The spirals should be denser and hotter  than the rest of the disk. The inner parts of the disk should be optically thin in the continuum or be totally void. 
A synthetic Doppler map constructed using  this model of the  accretion disk  can also  reproduce the observed maps, as  was recently shown  in \citet[Fig.4]{2010ApJ...711..389A} .

\begin{acknowledgements}
The authors greatly appreciate discussions with Michele Montgomery and her suggestions, which helped to improve the manuscript significantly. We are also thankful  to the anonymous referee for constructive comments.   SZ and  GT acknowledge PAPIIT grants IN-109209/IN-103912  and CONACyT grants 34521-E; 151858 for resources provided towards this research.

\end{acknowledgements}

\end{document}